# Mobility Behaviors Shift Disparity in Flood Exposure in U.S. Population Groups

Bo Li [a], Chao Fan [a,b,] *, Yu-Heng Chien [c], Chia-Wei Hsu [a], Ali Mostafavi [a]

[a] UrbanResilience.AI Lab, Zachry Department of Civil and Environmental Engineering, Texas A&M University, College Station, TX, 77843, USA   Email: boli@tamu.edu

[b] School of Civil and Environmental Engineering and Earth Sciences, Clemson University, Clemson, SC, 29634, USA   Email: cfan@clemson.edu

[c] Department of Industrial and Systems Engineering, Texas A&M University, College Station, TX, 77843, USA

**Acknowledgement**

This material is based in part upon work supported by the National Science Foundation under Grant CMMI-1846069 (CAREER). The authors also would like to acknowledge the data support from Spectus. Any opinions, findings, conclusions, or recommendations expressed in this material are those of the authors and do not necessarily reflect the views of the National Science Foundation or Spectus.

**Declaration of interest**

None.

# Mobility Behaviors Shift Disparity in Flood Exposure in U.S. Population Groups

**Abstract:** The current characterization of flood exposure is largely based on residential location of populations; however, location of residence only partially captures the extent to which populations are exposed to flood hazards. An important, though yet under-recognized, aspect of flood exposure is associated with human mobility patterns and population visitation to places located in flood prone areas. In this study, we analyzed large-scale, high-resolution location-intelligence data collected from anonymous mobile phone users to characterize human mobility patterns and the resulting flood exposure in coastal counties of the United States. We developed the metric of mobility-based exposure based on dwell time in places located in the 100-year floodplain. The results of examining the extent of mobility-based flood exposure across demographic groups reveal a significant disparity across race, income, and education level groups. The results show that Black and Asian, economically disadvantaged, and undereducated populations in US coastal cities are disproportionally exposed to flood due to their daily mobility activities, indicating a pattern contrary to that of residential flood exposure. The results suggest that mobility behaviors play an important role in extending flood exposure reach disproportionally among different socio-demographic groups. The results highlight that urban flood risk assessments should not only focus on the level of flood exposure to residences, but also should consider mobility-based exposure to better learn the disparities in flood exposure among social groups. Mobility-based flood exposure provides a new perspective regarding the extent to which floods could disrupt people's life activities and enable a better characterization of disparity in populations' exposure to flood hazards beyond their place of residence. The findings of this study have important implications for urban planners, flood managers, and city officials in terms of accounting for mobility-based flood exposure in flood risk management plans and actions.

1. **Introduction**

Cities around the world are experiencing significant increase in flooding risk caused by climate change and relative sea level rise (Arns et al., 2020; Milly, Wetherald, Dunne, & Delworth, 2002). Flood events have affected more than 2 billion people worldwide from 1998 through 2017, and even worse, the intensity and frequency of floods are increasing (WHO, 2023). Research revealed a notable risk of over 20% in the global population facing the potential risk of floods (Tellman et al., 2021). Understanding the extent and disparities in flood exposure plays a key role in formulating equitable flood risk management plans and policies. The standard approach to flood exposure assessment focuses on place of residence as the location from which the extent and disparities of hazard exposure are quantified. This approach is based on overlaying flood hazard map or floodplain areas with population distribution information to estimate flood exposure (Boulange, Hanasaki, Yamazaki, & Pokhrel, 2021; Chakraborty et al., 2022; Mohanty & Simonovic, 2021; Qiang, 2019). Studies focusing primarily on residential locations assume that the impact of floods on people is solely based on damage to their residence. While residential damage is the primary way floods affect populations, floods also disrupt life activities. Populations spend much of their time away from their residence, for purposes of work, education, recreation, and other life activities. Considering only "residential night-time population counts" may lead to inefficient understanding of the actually number of people who are at risk from flooding (Smith, Martin, & Cockings, 2016).

An important, and yet under-recognized, aspect of flood hazard exposure is the consideration of disruptions in life activities of people. People's life activities depend on community facilities (such as schools, grocery stores, and workplaces) whose inundation would cause significant social and economic impacts. Hence, it is essential to examine population flood exposure beyond the place of residence to better characterize flood exposure and disparities. In fact, how human dynamics shape hazard exposure disparity has emerged as a topic of considerable interest and significance among researchers. The extent to which human mobility exacerbate exposure has been studied in various the context of environmental hazards, such as air pollution, earthquake, and flooding. Early studies attempted to tackle the problem by differentiate night and daytime-specific population densities to indicate population dynamics into seismic hazard exposure assessment (Freire & Aubrecht, 2012). Reis et al. (2018) compared residence-and workplace-based population-level exposure to air pollution. Zhu et al. (2019) proposed an agent-based model to simulate dynamic flood exposure by taking residents' travel behavior into consideration, thus uncovering the variation of flood risk due to dynamic population distribution and response behaviors. While the early explorations realized the role of human dynamics and devoted efforts to overcome the limitation of relying solely on residential information, they all encountered challenges in procuring accurate, fine-scale human mobility data empirically.

More recently, the prevalence of large-scale high-resolution mobility dataset gathered from mobile device location records provides researchers with massive opportunities of observing human dynamics from a large spatial-temporal scale. GPS locations from opted-in anonymized users and the associated time information are used to extract human visitation information, which may include locations and durations. Such data is able to provide researchers with a massive number of samples in a rapid manner, thus facilitating the observation, estimation, and modelling of human mobility dynamics at a high spatial-temporal resolution (Yabe, Jones, Rao, Gonzalez, & Ukkusuri, 2022). Moreover, mobility data can be combined with socio-demographic information to provide new insights into equality and justice in various urban contexts, such as environmental issues, disaster response, and pandemic monitoring. Examples of the related research include capturing race and wealth disparities in disaster evacuation patterns (Deng et al., 2021), examining experienced income segregation related to human mobility behavior (Moro, Calacci, Dong, & Pentland, 2021), revealing socioeconomical and racial disparities in community resilience by analyzing changes in mobility behavior before, during and after a hurricane (Boulange et al., 2021), discovering income disparities in terms of mobility associated exposure to air pollution (Fan, Chien, & Mostafavi, 2022), and investigating mobility flexibility among difference income groups during COVID-19 pandemic (Iio, Guo, Kong, Rees, & Wang, 2021). The findings of these recent studies show that the examination of fine-grained and large-scale human mobility data at their intersection with population hazard exposure could provide novel insights regarding ways human dynamics shape hazard exposure.

For this purpose, this study examines a novel mobility-based approach to flood exposure based on analyzing large-scale high-resolution location-intelligence data collected from deidentified mobile phone users to characterize human mobility pattern and the resulting flood exposure in coastal counties of the United States. To assess mobility-based flood exposure in this study, a particular attention is paid to disparities in flood hazard exposure among sub-populations. The uneven distribution of flood exposure among population groups has been identified by many empirical studies from various geographic regions. For example, Qiang (2019) explored the socio-economic disparities in subpopulations' exposure to flood hazard and came to the conclusion that economically disadvantaged population in the United States are more likely to reside in flood zones at a national scale in the U.S. Tate, Rahman, Emrich, and Sampson (2021) found that the adverse effects of flood risk are disproportionately amplified for socially vulnerable populations; Rentschler, Salhab, and Jafino (2022) found the that nearly 90% of the flood-exposed population reside in low- and medium- income countries worldwide, indicating the imbalanced relationship between flood exposure and poverty. Sanders et al. (2022) uncovered the large and inequitable flood risk exposure in Los Angeles, which is disproportionately higher for non-Hispanic Black and other disadvantaged population groups. Chen et al. (2022) also examined urban-rural exposure to flooding at city scale and came to the conclusion that gender, education level and economic development level are the drivers to expose specific populations to higher flood risk. The uneven distribution of flood exposure usually poses larger threats to economically disadvantaged groups and race minorities, who are among the most vulnerable sub-populations. Despite known disparities regarding to residential-based flood exposure, little is known about whether disparities exist, and if so, the extent of disparities when taking human mobility behaviors into consideration of flood exposure. Accordingly, in this study, we mainly address two research questions: (1) What is the extent of disparity related in *mobility-based flood exposure* across different sub-populations? and (2) To what extent do *mobility activities change the disparity in flood exposure compared with residential flood exposure?*

To address these research questions, we examined the dynamics of human activities from a large geolocation dataset collected from mobile phones, and we integrated 100-year floodplain of coastal cities across the Contiguous United States (CONUS) to develop a human mobility-based flood exposure metric. Additional socio-demographic information was examined to characterize mobility-based exposure disparity and compared with the residential flood exposure disparity. This study and findings offer a deeper understanding of flood exposure by examining human network dynamics. Also, the findings provide empirical evidence regarding disparities in mobility-based flood exposure among different population groups. The finding show that human mobility extends flood-hazard exposure disproportionately to vulnerable populations including racial minorities, economically disadvantaged, and undereducated population groups. The findings of this study offer an alternative lens through which to view flood exposure inequality associated with the different mobility characteristics, which is unrecognized from the standard residential exposure perspective. The results also reveal the positive relationship between residential and mobility-based flood exposure disparity, implying the cooccurrence and reinforcement of the two disparities. The characterization of mobility-based flood exposure offers a fresh viewpoint on how extensively floods can interrupt daily life activities. This perspective also allows for a more nuanced understanding of the varying levels of flood hazard exposure across different populations, based on their mobility characteristics other than location of residence. The results of this research carry significant weight for urban planners, flood managers, and municipal officials. They underscore the importance of incorporating mobility-based flood exposure considerations into flood risk management strategies and measures.

## 2. Methods

### 2.1 Study context

The research region of this study encompasses the coastal countries in the CONUS. The list of coastline counties was retrieved from the US Census Bureau. The study areas are spatially distributed across three coastline regions: the Atlantic, Gulf of Mexico, and Pacific Regions (Bureau, 2019). Counties belonging to Alaska and Hawaii were excluded from the list. The remaining counties are located in 20 states. Roughly 40% of the US population live in the coastal counties, and the gross domestic product of coastal regions as a unit is ranked as third highest in the world (NOAA, 2018). The proneness to flood hazard, together with the huge population and economy size made this area a suitable study region for flood exposure characterization.

### 2.2 Mobility-based flood exposure framework

In this study, mobility-based flood exposure is calculated based on the amount of time (dwell time) individuals spend in places located in flood plains (Fig.1). The background in Fig.1 shows the estimated 100-year floodplain map, where the blue blocks highlight areas estimated to be exposed to a 100-year flood.

In this study, we refer to human mobility as macroscopic and short-range movement behaviors during which visitations to places occur. The reasons for the movements can be diverse, such as commuting, work, and performing social and leisure activities. Since the mobility patterns of a large population are regular and periodic (Barbosa et al., 2018), examining a period of mobility behaviors can provide a general basis for mobility-based flood exposure of population. Accordingly, using fine-grained, large-scale location-based data, we calculate the dwell time of populations at places located in flood prone areas in calculating mobility-based flood exposure. The greater the dwell time in flood prone areas, the greater the mobility-based flood exposure for a particular population. Similar with the studies on residential flood exposure, the extent of mobility-based flood exposure could vary across different socio-demographic groups, and thus, we examine the extent of disparities in mobility-based flood exposure and compare those with residential flood exposure. Residential flood exposure can be obtained by integrating geolocations of residence and floodplain map information. The shift of focus from the standard residential exposure assessment to mobility-based flood exposure reveals new insights regarding the extent of flood exposure inequality.

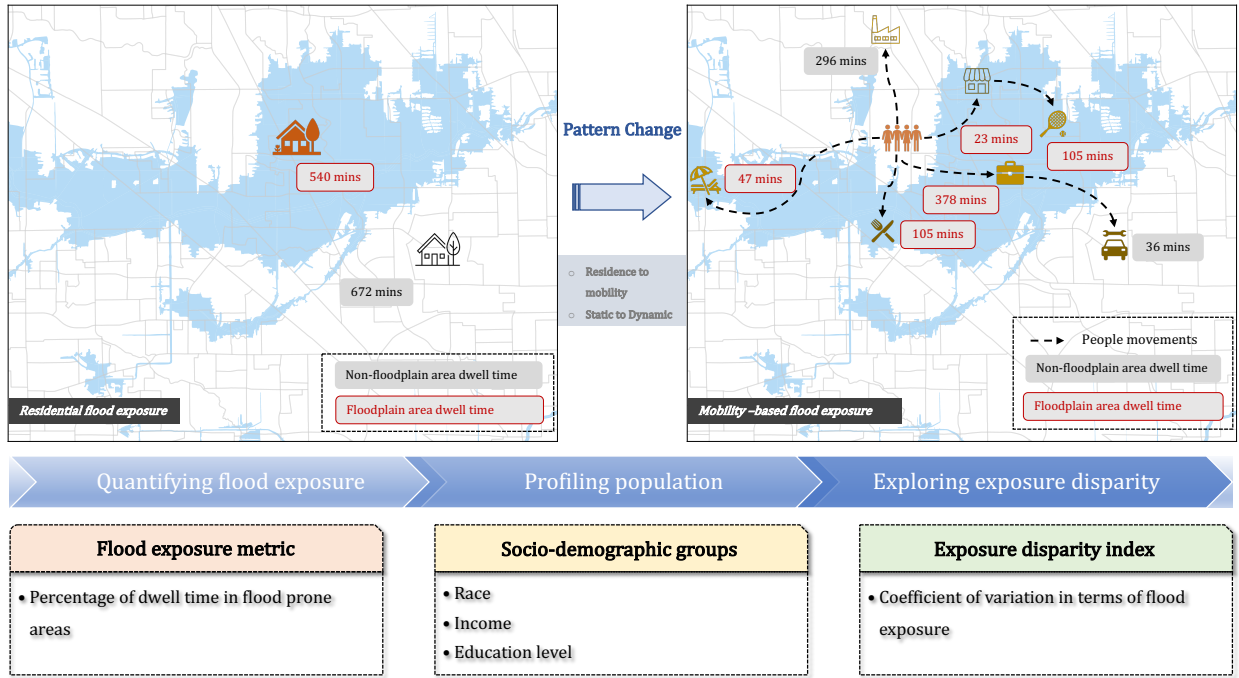

**Fig. 1 Mobility-based flood exposure and disparity.** The schematic illustrates the general framework of this paper. The flood exposure metric was developed by quantifying the dwell time in flood prone areas. Residential flood exposure examines whether individuals reside in the floodplain, while mobility-based flood exposure examines whether the places they visit are flood prone. Then socio-demographic information was integrated to determine the exposure disparities among specific population groups.

The overall analysis procedure comprises the following steps: first, we extracted residential information and human visitation records from location intelligence at a fine resolution, then overlaid floodplain maps of the studied area to identify flood prone areas. Dwell time in the flood prone areas was then calculated for individual users, and accordingly, we specified residential and mobility-based flood exposure. Finally, the extent of flood exposure inequality was computed for different socio-demographic groups across the study areas.

**2.3 Identifying residential areas and extracting dwell times from location intelligence data**

This study analyzed human mobility patterns at the census block group (CBG) level, which is a rather fine geographical level for which the US Census Bureau provide demographic sample data. Location intelligence data from Spectus (formerly known as Cuebiq) were used to extract population residential area (i.e., home CBG) and human visitation patterns. Spectus collected high-resolution data from anonymized smart phone users who opted in the relevant applications. It could support the need for data precision by providing device ID with corresponding home CBG ID, which we interpreted to be users at their residential CBGs. In this study, we only considered intra-county CBG visits (and excluded cross-county movements), since visits in this scale can best represent human's daily lifestyles. The time span we selected is the 7-day period from April 1through 7, 2019, during which no major event or holidays occured, thus could reflect the normal and steady state of human mobility. The research period contains both weekdays and weekends to capture the effects of weekday and weekend travel patterns. Visits were extracted from the geographically labelled pins with time stamps provided by Spectus. During the study period, each device ID has a record with latitude and longitude, and dwell time at different locations. The records for each device ID were compared with the corresponding home CBG information to determine whether the visits were within the residential CBG or outside the residential area. If the geographical coordinates of the records for a device were not located in the home CBG, and dwell time locations were in 100-year flood plains, then the dwell time would be summed up as dwell time in flood plains due to mobility. Similarly, the dwell time of a device within the home CBG is aggregated as residential dwell time.

**2.4 Identifying flood prone areas**

To determine whether a CBG is flood-prone area, this study adopted 100-year flood plain data provided by the US Federal Emergency Management Agency (FEMA, 2023). Estimated from models considering rainfall and hydraulic statistics, floodplain shows areas with certain probabilities of being flooded. One hundred-year floodplain indicates a 1% chance of inundation during the flood event every year, which is a longstanding and important reference for delineating flood risk (Blessing, Sebastian, & Brody, 2017). This study follows the common practice in flood exposure studies to adopt floodplain to determine areas which are at a risk of flooding. The 100-year floodplain was overlaid with coastal county shapefiles obtained from the US Census Bureau, and the overlapped areas were identified as areas with flood hazard exposure.

**2.5 Measuring flood exposure associated with human mobility**

We defined the flood exposure metric as the proportion of dwell time people spent within areas at a risk of flooding within a specified week. Specifically, we assumed devices' records in a CBG captured by Spectus are representative of the population of the CBG. We first determined if the stop points of users are in the flood prone areas to categorize the flood exposure. Then, the residential flood exposure was calculated by aggregating the dwell time of all stops of a user within the home CBG. The mobility-based flood exposure was computed based on the aggregated dwell time of all stops outside the home CBG. Total flood exposure is measured by the proportion of the time people spent in a week at residences or visited places that are in flood plains. Accordingly, the mobility-based flood exposure $e_m$ of a CBG $n$ is given by:

$$e_{mn} = \frac{\sum t_i}{\sum T_i} \tag{1}$$

where $t_i$ denotes the time device $i$ spent at places with flood exposure other than its home CBG in a week, and $T_i$ denotes the total time of a week, which is 10080 minutes. As such, we calculated all mobility-based flood exposure of users in CBGs within the study areas.

Next, to examine the possible distinctions between residential and mobility-based flood exposure, we computed flood exposure based on the residential information. Residential flood exposure is calculated based on the dwell time in home CBGs in flood prone areas. The residential flood exposure $e_r$ of a CBG $n$ is given by:

$$e_{rn} = \frac{\sum p_i}{\sum T_i} \qquad (2)$$

where $p_i$ denotes the time device i spent at its home CBG *n* in a week, and $T_i$ still denotes the total time of a week. Based on the definition, flood exposure is a unitless indicator, whose between 0 and 1. A direct observation from the formula reveals that the residential flood exposure of people living outside flood-prone areas is automatically 0, which reflects the static nature of residential flood exposure.

**2.6 Profiling exposed population groups**

Socio-demographic data were collected to characterize disparities in both residential and mobility-based flood exposure. We obtained data from the 2019 5-year American Community Survey at the CBG level (CensusBureau, 2022). Race, income, and education level are the three variables examined in this study for evaluation of flood exposure disparities, since the literature showed that these variables played a vital role in deciding vulnerability and differentiating hazard exposure (e.g.(Coleman et al., 2023; Forrest, Trell, & Woltjer, 2020; Sanders et al., 2022; Smiley et al., 2022)).. To categorize the race of each CBGs, we first computed the average proportion of the people in each race. The race of a CBG is determined if the proportion of people of a particular race is higher than the average proportion for a race among all coastal CBGs selected in this study. Similarly, we also calculated the average of the median household income of all selected CBGs. CBGs are categorized as high-income if their medium household income is higher than the average of all coastal CBGs selected in this study. For education, we employed educational attainment for the population 25 years and over to determine the education level of CBGs. CBGs having a higher proportion of people with lower educational attainment than the average were labelled as lower educational attainment, while the rest are considered as higher education.

## 2.7 Measuring flood exposure disparity

This study defined flood exposure disparity as the varying degree to which population groups are exposed to flood risk higher than a certain threshold. In this study, we determined the threshold as the global mean of flood exposure of the studied area, denoted as $T$. The proportion of those exceeding the threshold part compared to the threshold $T$ was defined as $q$.

Inspired by Jbaily et al. (2022), we adopted coefficient of variation (CoV) to indicate the degrees of flood exposure disparities at state level.. CoV is a statistic measuring the extent of variability in relation to the mean of the population whose formula is:

$$CoV = \frac{\sqrt{Var(q)}}{\mu(q)} \qquad (3)$$

where $Var$ is the variance of $q$, and $\mu$ is the mean of $q$. $CoV$ provides a standardized measure of variability by taking the mean of the dataset into consideration, such that the metric is more suitable for comparisons between data with different magnitudes and easier to interpret. For example, we can compute the flood exposure disparities for a certain state containing $CBG_1 \ldots CBG_n$: first compute $q_1, \ldots, q_n$ by subtracting the threshold $T$ for each CBG; then compute $Var(q_1, \ldots, q_n)$ and $\mu(q_1, \ldots, q_n)$ to obtain the value of $CoV$.

The measure $CoV$ is lower bounded by 0 representing complete equality, and thus a larger $CoV$ value means larger disparity. Based on the metric, we computed mobility-based and residential flood exposure disparity respectively.

## 3. Results

### 3.1 Disparities in mobility-based flood exposure

The results reveal the patterns of disparate flood exposure among different population groups. Fig. 2 illustrates the extent of mobility-based flood exposure among race groups. The majority of CBGs along the US coastal counties are categorized as majority White racial group, while majority Asian and Black groups represent a relatively small number of CBGs in certain areas.

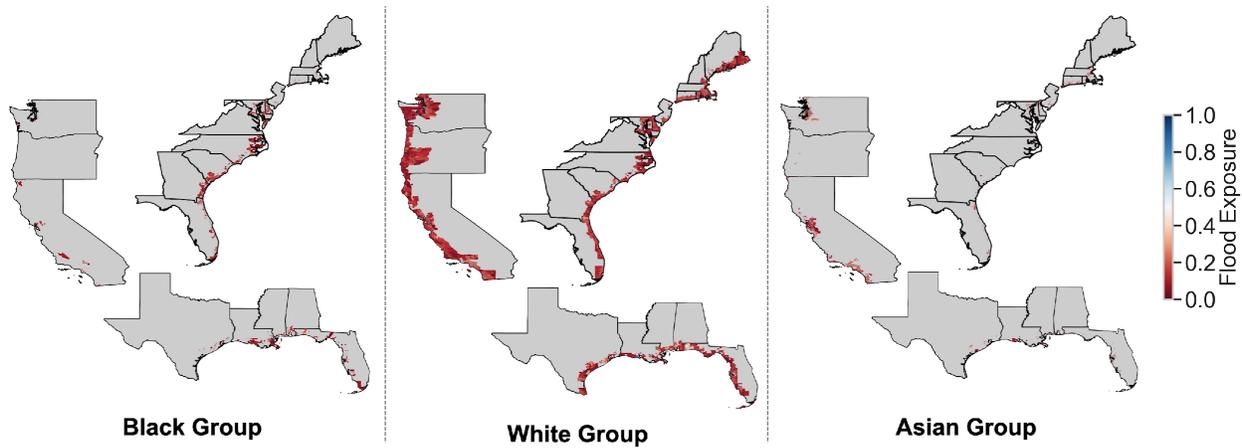

**Fig. 2 Spatial distribution of mobility-based flood exposure among race groups.** Mobility-based flood exposure is shown at the CBG level across coastlines in the United States. .. The magnitude of flood exposure is represented by a color bar. The White group has a wide spatial distribution along every coastline, while Asian and Black groups are scattered within several certain areas.

Fig. 3a shows the cumulative distribution function plot of mobility-based flood exposure for the three race groups. As shown, all the distributions of mobility-based flood exposure extent for the three race groups have a similar trend. Basically, the majority of the CBGs (more than 80%) experienced low levels of mobility-based flood exposure (less than 0.25). Then, the slope of the plots starts to increase exponentially, indicating a drastically greater mobility-based flood exposure for a smaller fraction of CBGs. The dramatic change indicates that a small proportion of the population (less than 20%) are exposed to a much greater level of flood exposure due to human mobility. The results reveal the long-tail distribution of mobility-based flood exposure among all population groups. In all population groups, a small proportion of populations endure the greatest mobility-based flood exposure, while a majority of populations have relatively smaller mobility-based flood exposure. Another important insight from these results is that, on average, the mobility-based flood exposure is around 0.15. This result suggests that the majority of populations have a mobility-based flood exposure of 0.15 irrespective of their residential flood exposure. The long-tail distribution of mobility-based flood exposure provides evidence of flood exposure inequality with each population group. Comparing the three plots, there is a perfect overlap of plots for Black and Asian groups, while the plot for the White group sees a slightly flatter increase, and the point where spike occurs come later. Specifically, only 5% of the White population is exposed to mobility-based flood exposure greater than 0.25, while the number is about 10% for Asian and Black groups. The differences in the plots indicate that generally the White group has less mobility-based flood exposure, and the proportion of population with extremely high mobility-based flood exposure is smaller in the White group. To further explore the potential inter-group difference, a t-test was performed between every combination of two race groups. The results confirmed the existence of group differences among all three group combinations with the significant level of 0.001. Fig. 3b showed that White group is the least exposed to flood, while the Asian group is most exposed , which indicates the disproportional distribution of mobility-based flood risk among race groups.

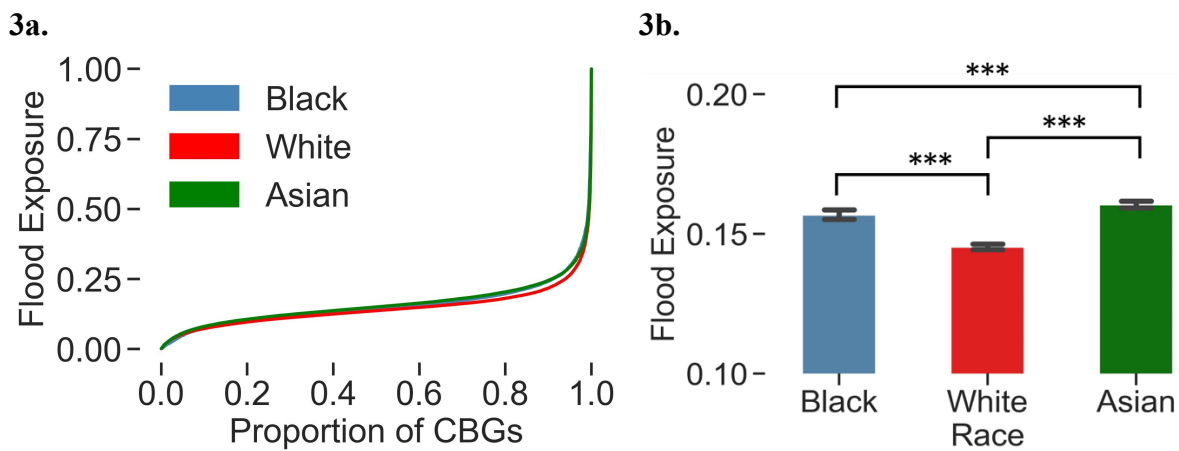

**Fig. 3 Distribution of mobility associated flood exposure among race groups**. **a.** Cumulative distribution function (CDF) plot for Asian, Black and White group. Note the blue curve is overlapped by the green curve. **b.** Boxplot showing the group difference of race with significance test. Note: *** $p < 0.001$.

We also performed similar analysis on income groups and education level groups. Fig. 4 mapped the spatial distribution of low-and high-income groups with their flood exposure level. The results show that more CBGs with high mobility-based flood exposure can be identified from low-income groups than those from high-income groups. The cumulative distribution function in Fig. 5a shows that the high-income group has lesser proportion of population with high mobility-based flood exposure. The t-test result in Fig. 5b shows significant level of group difference ($p < 0.001$) in terms of mobility-based flood exposure between the two income groups. Low-income groups are more at risk to mobility-based flood exposure than high-income groups.

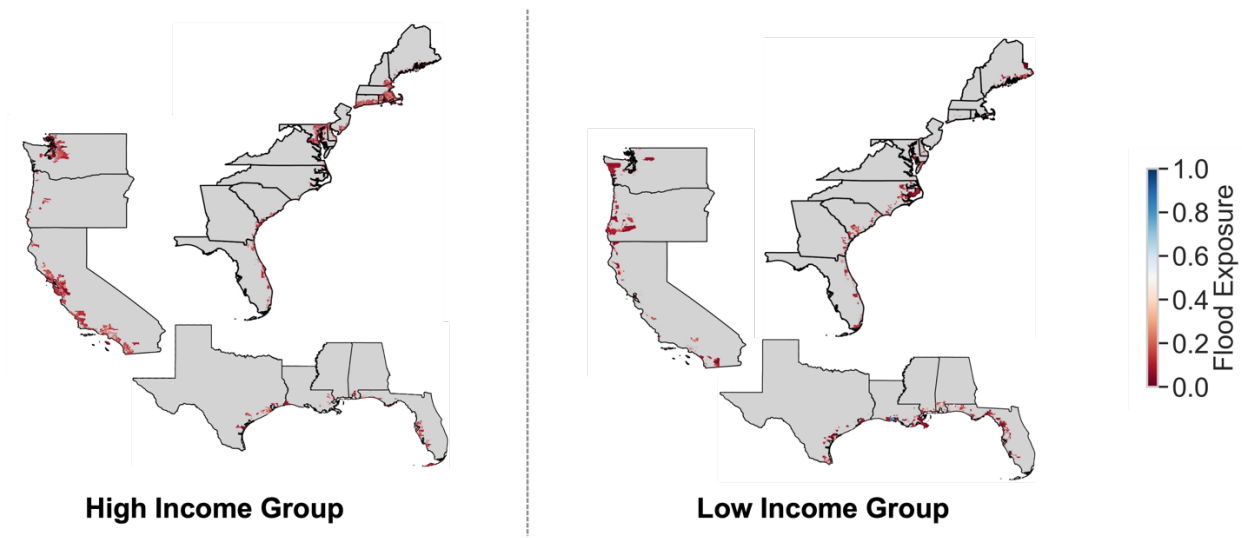

**Fig. 4. Spatial distribution of mobility associated flood exposure among income groups.** Mobility-based flood exposure is showed at CBG level across coastlines in the United States.. The magnitude of flood exposure is represented by a color bar.

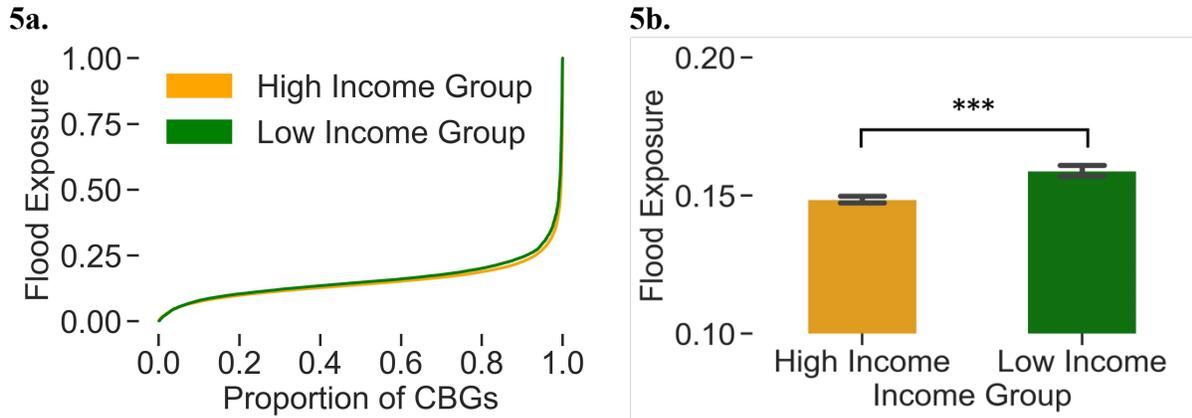

**Fig. 5 Distribution of mobility-based flood exposure among income groups. a.** Cumulative distribution function plot for high- and low- income group; **b.** Boxplot showing the group difference of income with significance test. Note: *** $p<0.001$.

A similar pattern can be seen for education level groups (Fig. 6 and Fig.7) that population group with a higher education level has lower proportion of people with high mobility-based flood exposure, and the average level of mobility-based flood exposure is significantly ($p < 0.001$) lower than population group with lower educational attainment.

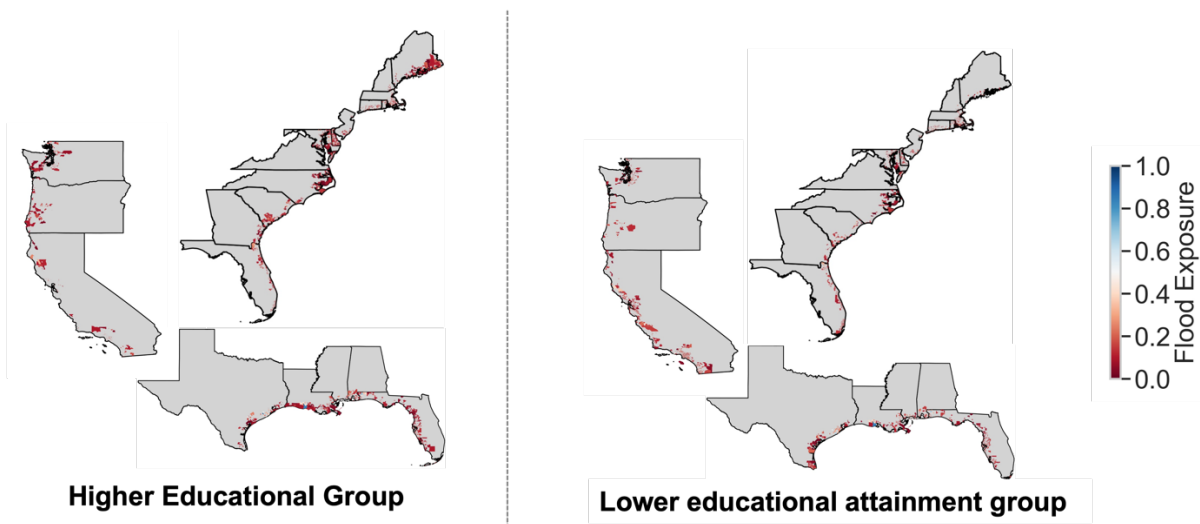

**Fig. 6. Spatial distribution of mobility-based flood exposure among education level groups**. Mobility-based flood exposure is shown at the CBG level across coastlines. The magnitude of flood exposure is represented by a color bar.

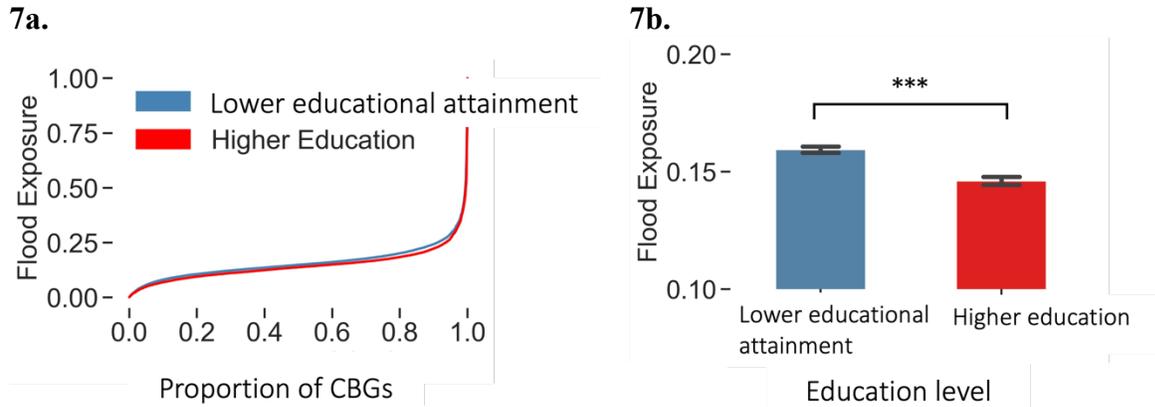

**Fig. 7 Distribution of mobility associated flood exposure among educational level groups**. **a.** Cumulative distribution function plot for lower educational attainment and higher education group; **b.** Boxplot showing the group difference of education level with significance test. Note: *** $p<0.001$.

### 3.2 Mobility exacerbates flood exposure disparity patterns

In the next step, we examined the extent to which mobility would exacerbate flood exposure compared with the standard residential flood exposure disparity pattern. Fig. 8 shows the two forms of flood exposure level among all CBGs in the research area. Generally, residential exposure has a higher level than mobility-based exposure, indicating that the residential flood exposure still poses a greater threat to the public. However, studying mobility-based exposure provides a more complete picture of the flood exposure risk by providing a complementary perspective. For example, people may live in low flood risk places, while they may travel to places with high flood risk areas, from which they may still suffer.

Fig. 9a displays the complementary cumulative probability density function (CCDF) of residential and mobility-based flood exposure, the y axis of which denotes the probability that flood exposure exceeds the certain threshold. The plot of residential flood exposure has a gradual decay initially, with the probability of flood exposure larger than 0.4 being approximately 80%. After that the curve shows a rapid decay until the residential exposure is 0.6 with the probability of 15%. This result suggests that a large proportion of the population has a moderate level of residential flood exposure, while a smaller proportion has a high level of exposure. The S-shape of the CCDF for the residential flood exposure suggest a logistic distribution pattern in which residential flood exposure values are symmetrical around the mean. In contrast, the plot of mobility-based flood exposure shows a long-tail distribution which decays rather sharply from the beginning until the probability that mobility-based flood exposure exceeds 0.25 is 10%, which echoed the observation that mobility-based flood exposure is on average less than the residential flood exposure. The logistic distribution of residential flood exposure implies a different kind of inequality, with less weight given to extreme flood exposure values. On the other hand, the long-tail distribution of mobility-based flood exposure suggests a more extensive inequality compared with residential flood exposure.

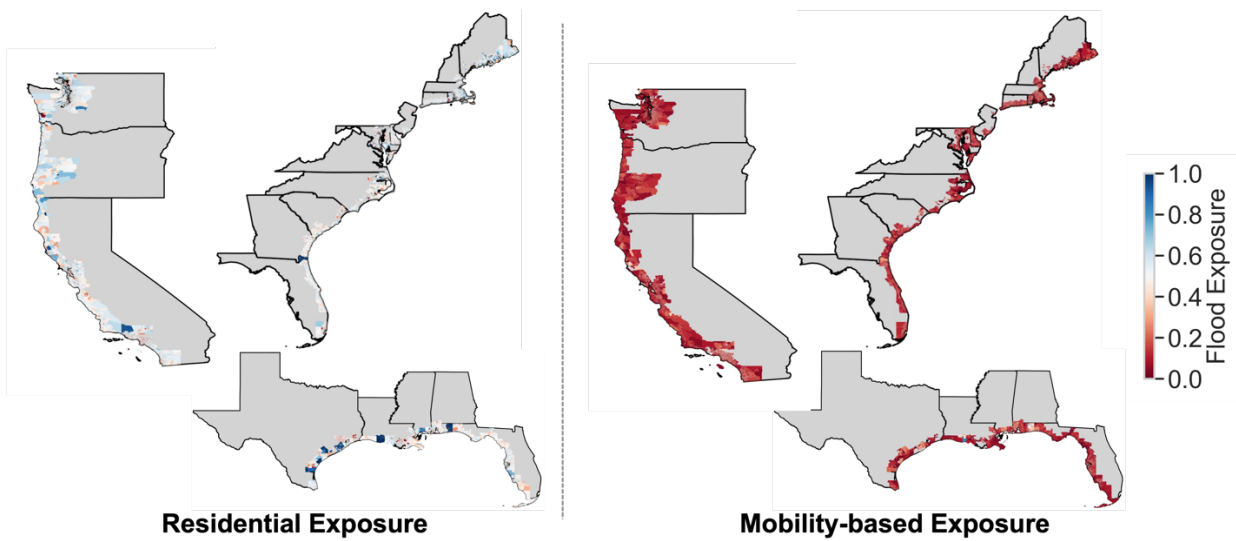

**Fig. 8 Geographical mapping of residential and mobility-based exposure**. Two forms of flood exposure are shown at the CBG level across coastlines in the United States. The magnitude of flood exposure is represented by a color bar.

Fig 9b, 9c, and 9d depict the comparisons between different population groups for residential flood exposure. For race group, the significance test only supports the existence of group difference between Black and White group ($p < 0.001$); Results for income and education level group show significant group differences between low-and high-income group, and lower educational attainment and higher education group, which is the same case as mobility-based exposure. Remarkably, when comparing the average values between the groups, the result shows a pattern

exact opposite from the case of mobility-based flood exposure: the Black racial, , low-income, and lower educational attainment groups have less residential flood exposure compared with the non-vulnerable groups. That means if solely relying on the residential flood exposure, the flood exposure disparity for vulnerable groups would be underestimated. The difference of social-demographic attributes may influence many life conditions, such as where they live, where they work, where they purchased supplies and where they go for life activities. Broadening the characterization of flood exposure from solely residence-based and considering mobility-based flood exposure uncovers more disparities which could not be observed before.

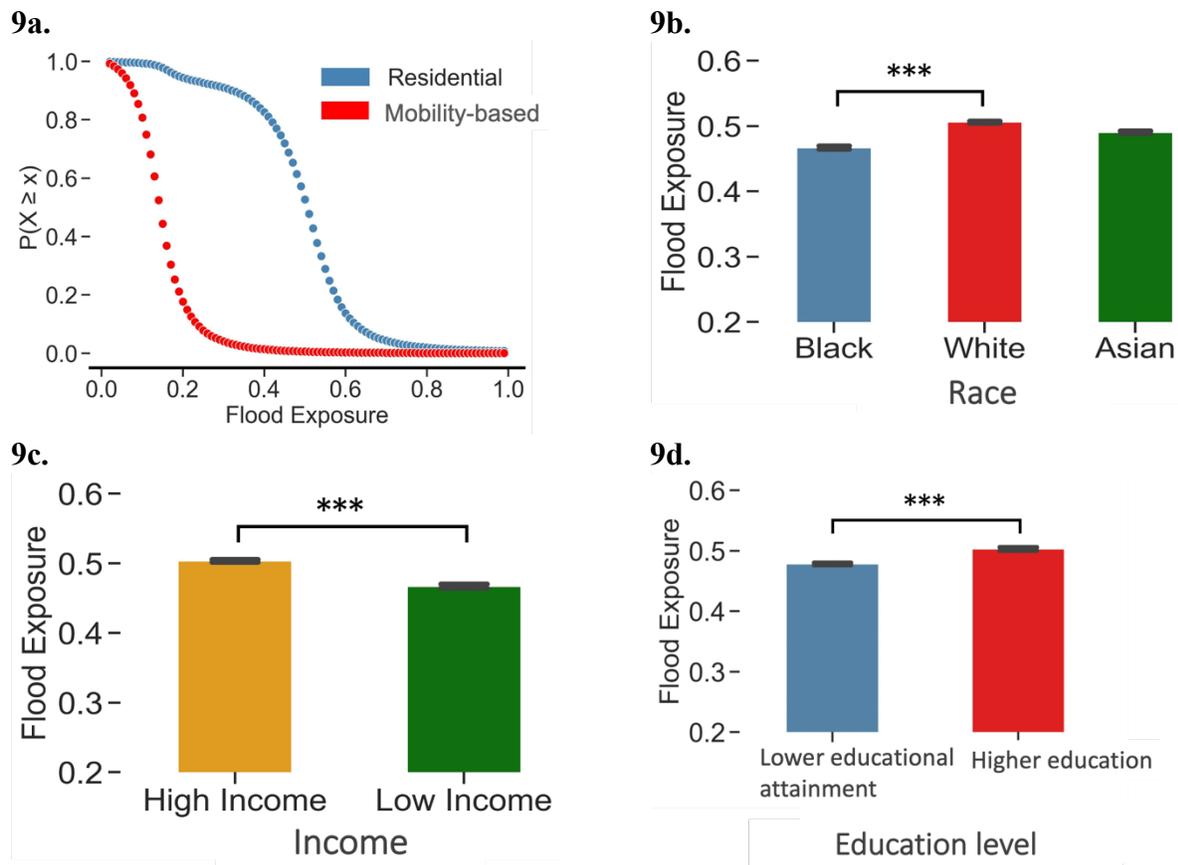

**Fig. 9 Distribution of residential flood exposure. a.** Complementary cumulative probability density function of residential and mobility-based flood exposure; **b, c, d.** Boxplot showing the group difference of residential flood exposure for race, income, and education level groups with significance test.

We also conducted state-level comparisons between residential exposure and mobility-based exposure. Fig. 10 displays the distribution of the two exposure metrics for all states. The distributions of residential exposure for Mississippi, Maine, Virginia, Texas are highly variable, while the mobility-based exposure distributions for those states consistently have a relatively wider range of values compared with other states, and vice versa for the states with low variability. The direct observation implies that there might exist a consistent relationship within states that high level of residential flood exposure is associated with high level of mobility-based flood exposure. To examine the association between disparities in residential versus mobility-based flood exposure at state level, we conducted a regression analysis between disparity index of residential and mobility-based flood exposure. Fig 11a. displays the scatter plot of the states, and a positive straight line fitted, confirming the positive association between disparities of residential and mobility-based flood exposure. Fig 11b. depicts the disparity for the two forms of flood exposures for all states. Although residential exposure generally has a higher level of disparity than mobility-based exposure, the pattern exists that places with greater disparity of residential flood exposure also have greater level of disparity in mobility-based flood exposure. This finding implies that once disparities are found to exist, the extent of disparity would be more severe than perceived because disparities result from two different sources may coexist and have a positive association with each other.

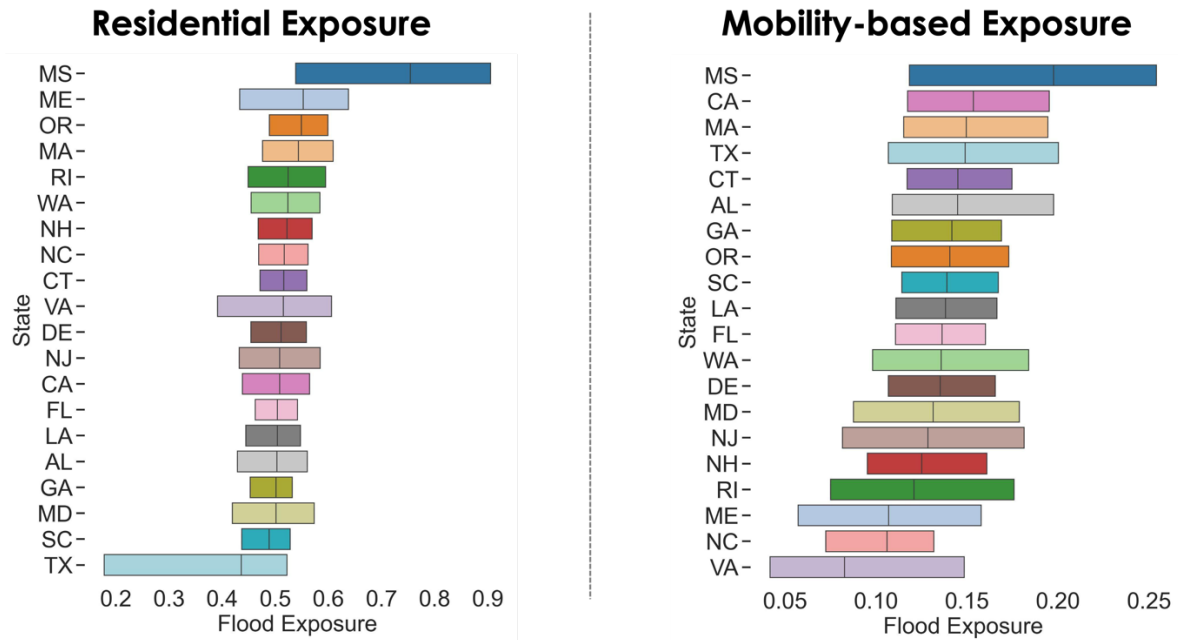

**Fig 10. Distribution of residential and mobility-based flood exposure for all states in the research area.**

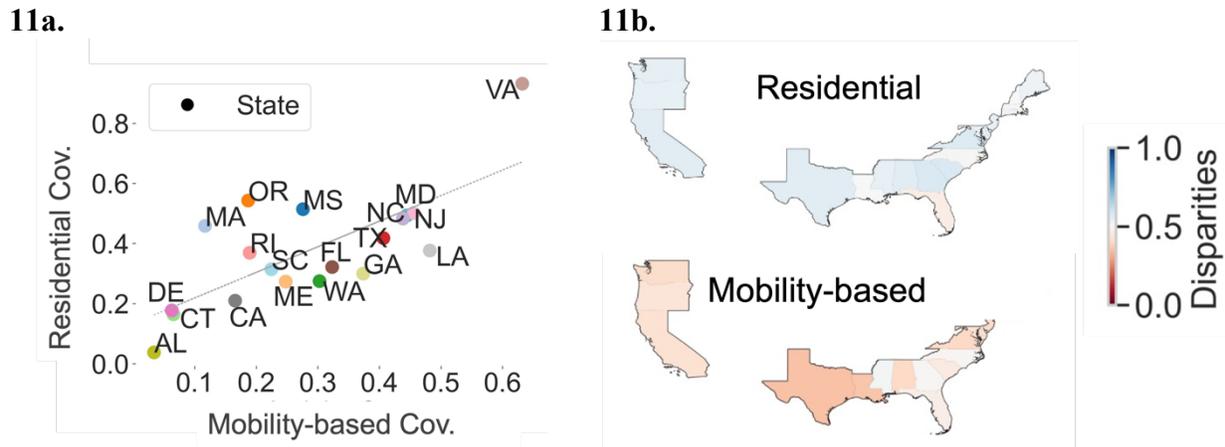

**Fig 11. Disparities for residential and mobility-based flood exposure risk**. **a.** Regression analysis of flood exposure disparity at state level; **b.** Geographical mapping of flood exposure disparity

## 4. Closing remarks

Although flood exposure disparity is a widely studied topic, the current characterization of flood exposure is primarily based on the location of residence, while overlooking the ways human dynamics alter flood exposure and disparities. The existing knowledge about whether and to what extent human mobility alters flood exposure is rather limited. To bridge the gap, this study examined the dwell time in places located outside the place of residence of individuals and created a novel mobility-based flood exposure metric. By leveraging large-scale, and fine-resolution mobility datasets, this study is enabled to capture human dynamics and quantify the time people spent within the flood plains. Further, the analysis compared the mobility-based flood exposure among various socio-demographic groups and found the presence of disparities among race, income, and education level groups.

The main findings of this study are fourfold. First, the mobility-based flood exposure follows a long-tail distribution suggesting a severe inequality. Also, mobility-based flood exposure is highly disproportional among the socio-economic population groups: white, high-income, and well-educated people have less exposure to flood based on their mobility activities compared with the vulnerable groups. This finding indicates that vulnerable population groups could endure greater social and economic impacts from a greater mobility-based flood exposure. Prior research revealed that different social groups experience various levels of flood risk by taking built-environment conditions of residence into consideration (e.g.(Chang et al., 2021; Chen et al., 2021)), while this study opens the venue for understanding flood exposure from the perspective of human mobility dynamics. Mobility-based flood exposure is shaped by the spatial distribution of flood hazards, facilities, and access, as well as lifestyle patterns. For example, prior studies reported the difference in travel distance and travel frequencies across income groups (Barbosa et al., 2021). These different mobility characteristics will further shape disparities in flood exposure among population groups. Second, the results revealed a different and more dire inequality in mobility-based flood exposure compared with residential flood exposure. The long-tail distribution of mobility-based flood exposure suggest a different kind of inequality compared with the logistic distribution of residential flood exposure. Third, the results indicate that Black, low income, and those with lower educational attainment have less residential flood exposure, which seems contradictory with some other studies. The inconsistency may result from the differences in study areas and the way residential flood exposure is calculated. Qiang (2019) found that the minority and disadvantaged groups are more inclined to live in flood prone areas in inland cities, while in coastal cities middle and upper-income groups occupy the waterfront properties and coastal areas because of their high values. Also, the results show that mobility-based exposures are completely flipped compared to residential exposure in terms of socio-demographic disparities. Vulnerable populations have a greater mobility-based flood exposure while non-vulnerable groups have a greater residential flood exposure. A greater mobility-based flood exposure for vulnerable groups imply that they would experience greater disruptions in their life activities during flood events. People spend much of their time away from residences for purposes of working, shopping for necessities, healthcare, entertainment, and so on. Residential flood exposure can be managed with property flood insurance which is affordable for non-vulnerable groups. However, mobility-based flood exposure could mean not being able to conduct daily life activities, which would have dire social, economic, and well-being impacts. The fourth finding is from the comparison among the states. The flood exposure disparity level of coastal areas among the different states is relatively consistent and proportional. This finding implies that states with higher residential flood exposure disparity would also have a greater mobility-based flood exposure disparity. The co-occurrence of the two types of flood exposure disparities exacerbates inequities.

As an early attempt to capture and quantify flood exposure based on human dynamics, the study has some limitations to be addressed in future research. First, the mobility-based flood exposure is calculated based on the dwell time in other CBGs outside their residential area. Hence, we are not able to differentiate the categories of places people visit outside their home CBGs, which impedes understanding the underlying activities that drive people to visit flood prone areas. Future research could build upon the current study by computing the dwell time to various categories of points of interest and perform more in-depth analyses. Second, although the 100-year floodplain is a longstanding and universally used metric to determine flood exposure in the United States, prior research e.g.((Highfield, Norman, & Brody, 2013)) has noted some of its drawbacks, including susceptibility to potential error upon implementation and, insufficiency to fully disclose the flood risk. Blessing et al. (2017) echoed the opinion that actual flood loss can occur outside the floodplain. Future studies can adopt other flood exposure models to mitigate this limitation. The method and metrics created in this study to calculate mobility-based flood exposure and flood exposure disparity can still be applied and are compatible with the improved models to more precisely characterize population exposure to floods.

**Data Availability**

The data that support the findings of this study are available from Spectus, but restrictions apply to the availability of these data, which were used under license for the current study. The data can be accessed upon request submitted on spectus.ai. Other data we use in this study are all publicly available.